\documentclass[conference]{IEEEtran}
\usepackage{cite}
\usepackage{amsmath,amssymb,amsfonts}
\usepackage{algorithmic}
\usepackage{graphicx}
\usepackage{textcomp}
\usepackage{xcolor}
\usepackage[hyphens]{url}
\def \sol {Q-BEEP}
\usepackage{algorithmic}
\usepackage{algorithm}
\usepackage{braket}
\usepackage{subfigure}
\usepackage{subfig}

\def\BibTeX{{\rm B\kern-.05em{\sc i\kern-.025em b}\kern-.08em
    T\kern-.1667em\lower.7ex\hbox{E}\kern-.125emX}}

\title{Q-BEEP: Quantum Bayesian Error Mitigation Employing Poisson Modeling over the Hamming Spectrum}
\author{\IEEEauthorblockN{Samuel Stein*, Nathan Wiebe${^\diamond}$*, Yufei Ding${^\dagger}$, James Ang*, Ang Li* }
\IEEEauthorblockA{\textit{*Pacific Northwest National Laboratory, Richland, Washington, USA}\\
\textit{${^\dagger}$University of California, Santa Barbara, California, USA}\\
\textit{${^\diamond}$University Of Toronto, Toronto, Ontario, Canada}\\
{\{samuel.stein, nathan.wiebe, ang, ang.li\}}@pnnl.gov, yufeiding@cs.ucsb.edu}}

\begin{document}
\maketitle

\begin{abstract}

The field of quantum computing has experienced a rapid expansion in recent years, with ongoing exploration of new technologies, a decrease in error rates, and a growth in the number of qubits available in quantum processors. However, near-term quantum algorithms are still unable to be induced without compounding consequential levels of noise, leading to non-trivial erroneous results. Quantum Error Correction and Quantum Error Mitigation are rapidly advancing areas of research in the quantum computing landscape, with a goal of reducing quantum errors. IBM has recently emphasized that Quantum Error Mitigation is the key to unlocking the full potential of quantum computing in a published article. A recent  work, namely HAMMER, demonstrated the existence of a latent structure regarding post-circuit induction errors when mapping to the Hamming spectrum. However, they assumed that errors occur solely in local clusters, whereas we observe that at higher average Hamming distances this structure falls away. Our study demonstrates that the correlated structure is not just limited to local patterns, but it also encompasses certain non-local clustering patterns that can be accurately characterized through a Poisson distribution model. This model takes into account the input circuit, the current state of the device, including calibration statistics, and the qubit topology. Using this quantum error characterizing model, we developed an iterative algorithm over the generated Bayesian network state-graph for post-induction error mitigation. Our Q-Beep approach delivers state-of-the-art results, thanks to its problem-aware modeling of the error distribution's underlying structure and the implementation of an Bayesian network state-graph. This has resulted in an increase of up to 234.6\% in circuit execution accuracy on Bernstein-Vazirani circuits and an average improvement of 71.0\% in the quality of QAOA solutions when tested on 16 real-world IBMQ quantum processors. For other benchmarks such as those in QASMBench, a fidelity improvement of up to 17.8\% is attained. Q-Beep is a light-weight post-processing technique that can be performed offline and remotely, making it a useful tool for quantum vendors to adopt and provide more reliable circuit induction results. Q-Beep will be released on GitHub.
\end{abstract}

\section{Introduction}
Quantum computing has drawn significant attention in recent years, with quantum computers and quantum related technologies being developed at unprecedented rates \cite{national2019quantum,bertels2021quantum,preskill2018quantum}. Quantum computing typically involves restricting a quantum mechanical system to a two-level system, and executing quantum operations on them to perform gate based computation. Quantum computing is poised to provide computational advantages that classical computing could never feasibly attain, with applications in domains such as quantum chemistry \cite{lanyon2010towards,fedorov2022vqe,cao2019quantum}, machine learning \cite{schuld2019quantum,stein2020qugan,broughton2020tensorflow}, arithmetic \cite{ruiz2017quantum}, optimizations \cite{wang2018quantum,farhi2016quantum,farhi2014quantum}, etc. Although algorithms that can provide computational speedups have been theorized, the current state of quantum computing suffers from substantial \textbf{size} and \textbf{noise} problems, rendering the present near-term noisy intermediate scale quantum (NISQ) processors being unable to offer computational advantages. 


Contemporary quantum computers have limited size, which physically limits the number of qubits available to an algorithm on a single quantum processor. Growing quantum processor sizes and algorithmic approaches to tackling this problem are actively being researched. Currently, the largest current quantum processor is IBM Osprey, comprising 433 qubits, and algorithmic scaling methodologies are being researched such as CutQC \cite{tang2021cutqc} or EQC \cite{stein2022eqc}, distributing one algorithm across multiple processors exploiting parallelism.

\begin{figure}
\centering
\begin{minipage}{0.35\linewidth}
\centering
\includegraphics[width=0.95\linewidth]{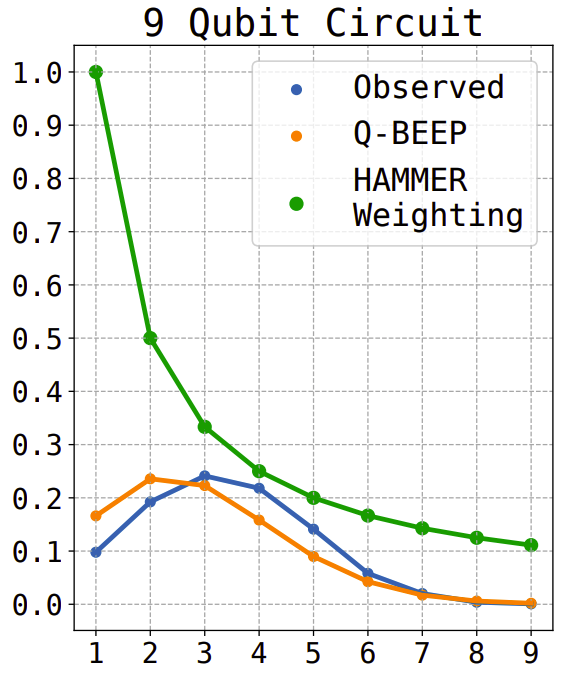}
\caption*{(a)}
\end{minipage}
\hfill
\begin{minipage}{0.6\linewidth}
\centering
\includegraphics[width=0.95\linewidth]{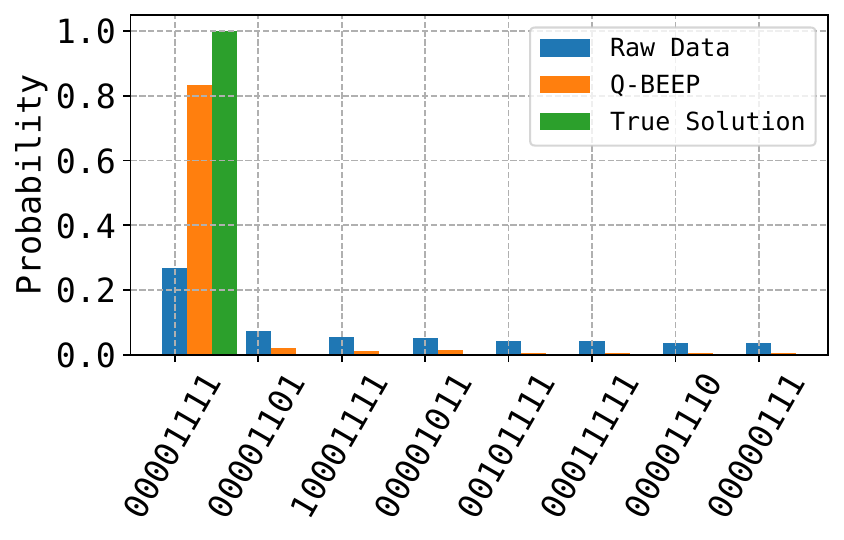}
\caption*{(b)}
\end{minipage}
\caption{(a) Example Hamming Spectrum where Q-BEEP is able to capture latent error structure, whereas HAMMER can not. (b) Quantum Error Mitigation for Bernstein Vazirani (BV) using Q-BEEP. Raw data is unprocessed data from an 8 qubit BV on a real quantum processor. The orange bar is the resultant probabilities through Q-BEEP. The green bar is the ideal observable bit-string probabilities.}
\label{fig:bv_error_mit}
\end{figure}

With respect to noise, there are a multitude of noise factors that impede sufficient performance on quantum processors. These noise factors originate from imperfect operations performed on quantum processors, state decay, spin-spin relaxation, state preparation and measurement errors, etc \cite{erhard2019characterizing,michielsen2017benchmarking}. These errors result in the bit-strings observed having a probability of being incorrect, and consequently lead to the induction of an algorithm containing both a mixture of correct and erroneous outputs. If the error rate is non-trivial, useful information is difficult to extract, leading to useless results. To mitigate this noise, there are two major approaches: \emph{quantum error correction} (QEC) and \emph{quantum error mitigation} (QEM). With QEC being too expensive in the near term, and IBM stating that QEM is the way to useful quantum computing \cite{ibm_2022}, we focus on QEM in this paper. 

QEM suppress the errors through repeated experiments and postprocessing of data. QEM algorithms are often based on structural indication such as symmetry \cite{cai2021quantum}, statistics \cite{strikis2021learning} and machine learning \cite{kim2020quantum}. Each of these works is driven by the motivation to understand and address the error structures observed across various scenarios. Earlier this year, Tannu et al. presented HAMMER \cite{tannu2022hammer}, stating that the erroneous measurement results are are not completely random; they observe that there exists a latent Hamming spectrum structure around the true solution. This observation motivates us to investigate the question:

\begin{center}
\textbf{\emph{What is this latent Hamming spectrum structure is, and how we can model this structure for Quantum Error Mitigation (QEM)?}}
\end{center}

In this paper, we find that the latent structure does not only follow the locality intuition, but can exhibit non-local clustering patterns. After examining the Index of Dispersion, a metric that measures the clustered nature of a dataset, across several observed hamming spectrums and various distribution models, we found that the Poisson distribution model is the most accurate in describing the complex, non-local clustered patterns observed. Employing the Poisson distribution structure, we propose a lambda estimation model that takes into consideration the input circuit, the hardware calibration statistics, and the qubit topology, to model the Hamming spectral errors as a function of both transpiled circuit and device. We propose an iterative graph-updating algorithm over the Bayesian network state-graph generated from this model. Through the model and the Bayesian approach, Q-BEEP can boost fidelity by up to 234.6\% on a \emph{Bernstein-Vazirani} (BV) circuit, or an 11.2$\times$ \emph{Probability-of-Success Trials} (PST)\footnote{Probability-of-success trials is defined later in Section IV-B} improvement.

We apply our corrective model to a diverse suite of applications. Our evaluation comprises analysis of \sol{} on (i) 1330 BV circuit inductions of width 5-15 qubits, across 8 different IBMQ machines, (ii) a large subset of QASMBench results \cite{li2020qasmbench}, comprising 224 circuits over 16 IBMQ machines containing algorithms such as Quantum Fourier Transform, Quantum Linear Solver and a Quantum Adder, improving the fidelity of past run circuits by up to 18.98\%, and an average of 6.67\%, and (iii) 340 \emph{Quantum Approximate Optimization Algorithm} (QAOA) solutions of varying p-values and problem graphs, sourced from Google's recent QAOA work \cite{harrigan2021quantum}, attaining an average Cost Ratio\footnote{Cost Ratio is defined later in Section IV-C} relative improvement of 1.71x compared to prior un-optimised solutions. Across our evaluation we provide consistent improvements, boosting domain-specific metrics and global fidelities.




This work makes the following contributions:
\begin{itemize}
    \item \textbf{Theory}: \sol{} presents a comprehensive examination of the latent structure of quantum errors in the Hamming spectrum of both trapped ion and superconducting systems. The study reveals the presence of non-local clustering patterns in the errors, which can be characterized using a Poisson distribution model. A novel method for characterizing these errors is proposed, incorporating circuit, hardware, and runtime features.
    \item \textbf{Technique}: \sol{} proposes an iterative Bayesian network update methodology to correct quantum errors based on the Hamming spectral errors predicted from the model. Given raw circuit induction results from a quantum device, Q-BEEP can adjust the bit-strings and their distribution to significantly boost fidelity (by up to 234.6\%) for general circuits without domain-specific implications or constraints. 
    \item \textbf{Demonstration}: \sol{} is comprehensively evaluated across over 1894 circuits from a diverse set of problems on 16 IBMQ NISQ devices, and provides state of the art performance across all categories, including fidelity and PST for BV, and CR for QAOA over prior art \cite{tannu2022hammer}.
\end{itemize}

\section{Background}
\subsection{Near-Term Quantum Computing}
In the present NISQ-era \cite{preskill2018quantum}, quantum computers are prohibited by their sizes, coherence times, gate fidelities, and a myriad of other computationally prohibitive factors. Quantum computers within the NISQ-era currently are dominated by two technologies, namely trapped-ion \cite{bruzewicz2019trapped} and superconducting \cite{clarke2008superconducting}.  Examples of these technologies are IonQ's 5-qubit trapped ion quantum computer, and IBM's quantum computers ranging in processor size from 1 to 433 qubits \cite{ibmquantum}. Alternative technologies have been proposed and implemented, such as liquid state nuclear magnetic resonance \cite{vandersypen2000liquid}, and free atom quantum devices \cite{ladd2010quantum}. Furthermore, with respect to the high-level structure of quantum computers, notions of the best structure continues to be researched. For example, the idea of distributed quantum computing using inter-quantum processor entanglement through inter-fridge \cite{sanz2018challenges} or intra-fridge \cite{usami2021photonic} communication is becoming more prevalent with protocols for communication being investigated \cite{haner2021distributed}. 

Gate based NISQ-era quantum computers are characterized by their physical qubit number, the low-level instruction set or basis gate set (e.g., CX for IBMQ), topology, and the runtime statistical calibration data. These numerical statistical properties characterize operation quality and are attained during benchmarking. Current performance and system size is prohibitive to inducing any useful algorithm attaining quantum supremacy. Attaining improvements in system performance and size, quantum error correcting codes, and improved quantum error mitigation techniques, are key to traverse this NISQ-era for real quantum advantage \cite{bravyi2022future}.

\subsection{Information Theory of Quantum Computing}

With quantum computers, a single algorithm induction results in one discrete result per pass. However, quantum computing results can change per pass, with probabilities according to the pre-measurement state. There is substantial variance in the world of quantum algorithms with respect to the diversity in system outputs. Certain algorithms aim to identify a unique output, such as the Quantum Adder, Bernstein Vazirani algorithm, or the Grover Search algorithm. Conversely, algorithms such as the Quantum Fourier Transform, Quantum Approximate Optimization Algorithm, and Quantum Phase Estimation showcase highly diverse outputs with no dominant single bit-string. In-between these algorithms exist algorithms that can contain multiple solutions comprising a subset of the entire $2^n$ solution space. The Shannon entropy theory can be leveraged to characterize the diversity of an algorithm’s observables. Higher-entropy algorithms contain more diverse outputs, while lower-entropy algorithms, indicates a sole fixed bit-string output.

\emph{Hamming distance} is a measure of similarity between two binary bit-strings, where the distance between them is described by $Ham(X,Y) = \sum_{i=0}^{len(X)}|X_i-Y_i|$. The Hamming distance is a metric that calculates the number of differences between each corresponding bit in two bit-strings, thus providing a notion of distance between solutions. Using Hamming distance, we can generate the \emph{Hamming Spectrum}, which represents a compact representation of the output probability distribution by bucketing each outcome into Hamming bins defined by distinct Hamming distance. 

Algorithm performance can be characterized through the use of Fidelity. Fidelity is commonly used in quantum computing to measure the distance between quantum states. Fidelity is computed via $F(\rho,\sigma) = (\sum_i^{2^n}\sqrt{\rho_i \sigma_i})^2$

With respect to the readout results on quantum computers, the results of quantum computers are usually interpreted as a pairs of bit-string to observation count. This high-level string representation of results inherently loses all spatial information within the Hamming spectrum. To capture the spatial nature of results, \emph{State Graphs}, whereby each observation bit string represents a node in a graph, can be adopted \cite{tannu2022hammer}. In this graph, vertices represent observations and the edges between them signify a connection between the two observations. An example is presented in the Q-BEEP architecture, Figure \ref{fig:section_guide}.

\subsection{Quantum Errors and Error Mitigation}

Quantum errors characterize one of the most prevalent issues in the NISQ era. Inducing a quantum algorithm on a NISQ hardware generates a mixture of both correct results, and errors. The ability to discern between the two is extremely challenging. As algorithms grow in complexity and size, the ability to successfully garner information from output distributions becomes exceedingly more crucial. 

Predominantly, Quantum errors are characterized by low qubit stability, operation infidelity, and state preparation and measurement (SPAM) errors. Quantum systems exhibit decoherence, which is the decay from a quantum state $\ket{\psi}$ to the ground state $\ket{0}$, as well as spin-spin relaxation, which characterizes the loss of phase-information in quantum systems. Gate infidelity and SPAM errors are the quoted success rates for each operation. These properties limit the depth of quantum circuits without introducing catastrophic error. 

Quantum error mitigation (QEM) approaches exist by means of pre- and post-circuit operations. For example, transpilation optimization can reduce global error rates through gate cancellation \cite{li2019tackling}. Pre-circuit induction transpilation has proven to provide substantial reductions in circuit complexity, with other approaches from the low level gate cancellation approach to dividing circuits into subsections and approximating unitary operators for sub-blocks \cite{patel2022quest}. These QEM techniques can provide relatively cheap error mitigation, as they operate on the classical side of quantum computing, and do not require tackling the systems' $2^n$ observables, or any hardware architectural change. 
 
Although algorithms and techniques continue to develop for error mitigation, quantum computing hopes to develop larger systems with improved system-wide performance statistics such as increasing T1/T2 times and improved gate fidelity. The combination of algorithmic error mitigation, and correction, in conjunction with improving quantum computer performance, allows for continued increasing capabilities of quantum processors.

\subsection{Quantum Error Structure}

Demonstrated in HAMMER, errors in resultant bit string distributions exhibit a clustered structure in the Hamming spectrum. This is empirically demonstrated through the notion that the expected Hamming distance of $\frac{n}{2}$, where n is the number of qubits in a system, is greater than the observed expected Hamming distance for many applications, hence indicating that errors are indeed clustered and localised. They go on to exploit this structure to provide a quantum error mitigating routine whereby the probabilities of each output are modified according to a function correlated to the inverse Hamming distance of other observed results. It is further demonstrated how varying degrees of entanglement do not disrupt this Hamming structure. As the fidelity of a system drops, the results approach being predominantly noise, which exhibits the largest expected Hamming distance of $\frac{n}{2}$. Given HAMMER is state-of-the-art and no alternative works are found, we target to compare directly with HAMMER in our evaluation.

\begin{figure*}[!ht]
    \centering
    \includegraphics[width=0.8\textwidth]{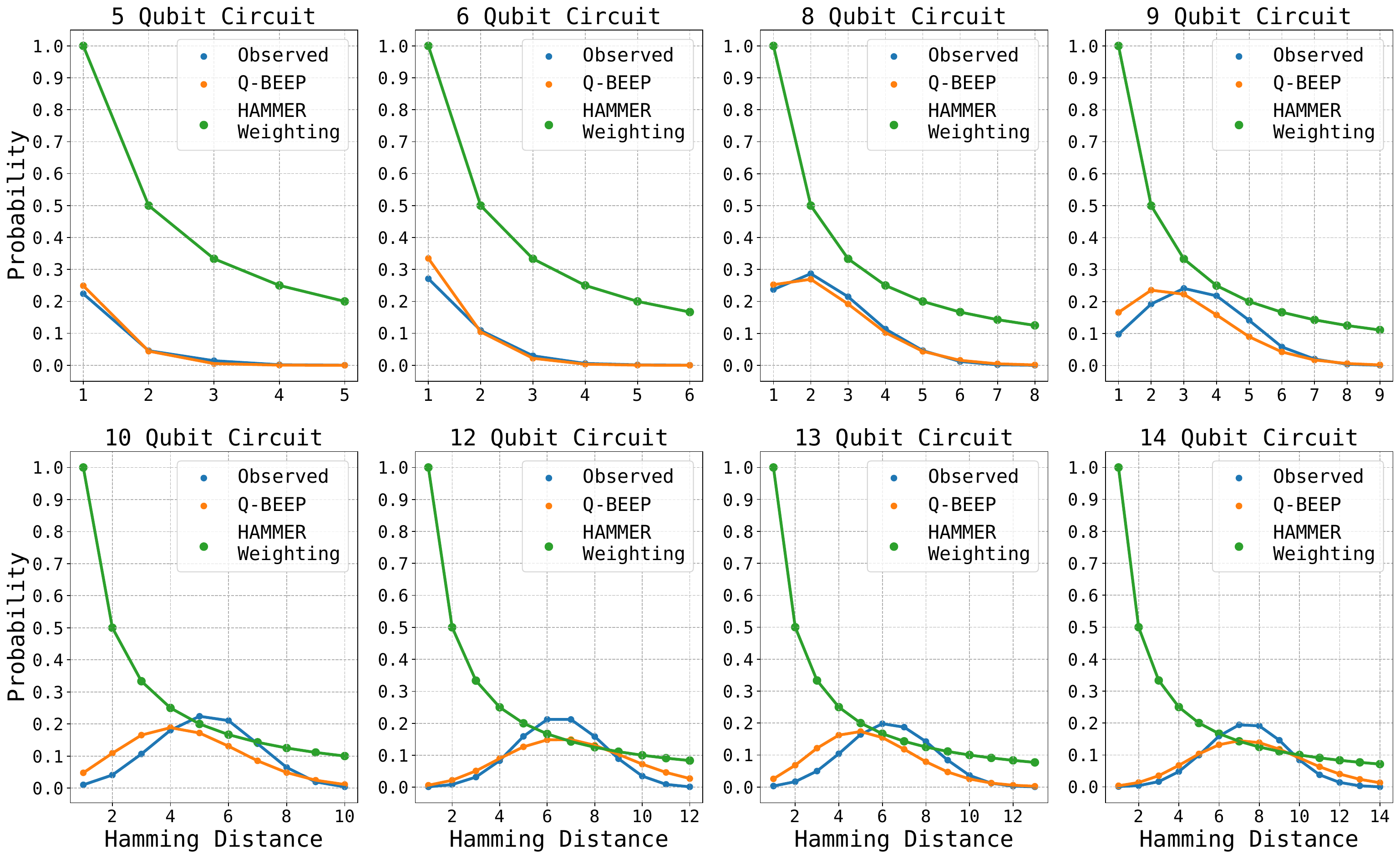}
    \caption{Sample comparison between observed probability distribution and Q-BEEP predictive probability distribution computed prior to the induction of BV circuits. Circuit qubit count labelled above subgraph, with Y-axis representing the probability of bit-string with respective Hamming distance observation, and X-axis the respective Hamming-distance. Each sub-graph is an independent experiment on an IBMQ device.}
    \label{fig:sample_distributions}
\end{figure*}

\begin{figure}
    \centering
    \includegraphics[width=0.38\textwidth]{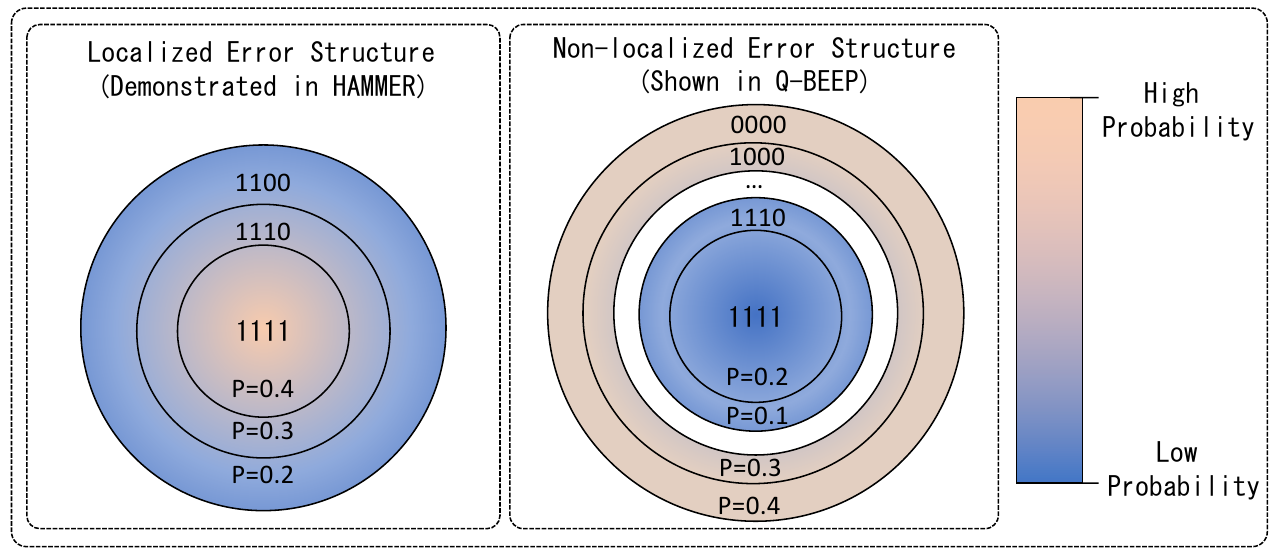}
    \caption{Hamming spectrum error structure where the left block shows the neighborhood structure in HAMMER, and the right block shows the distant neighborhood structure found by Q-BEEP, demonstrated in Figure~\ref{fig:sample_distributions}.}
    \label{fig:strucs}
\end{figure}

\section{Methodology}

\begin{figure*}
    \centering
    \subfigure[]{
    \includegraphics[width=0.3\textwidth]{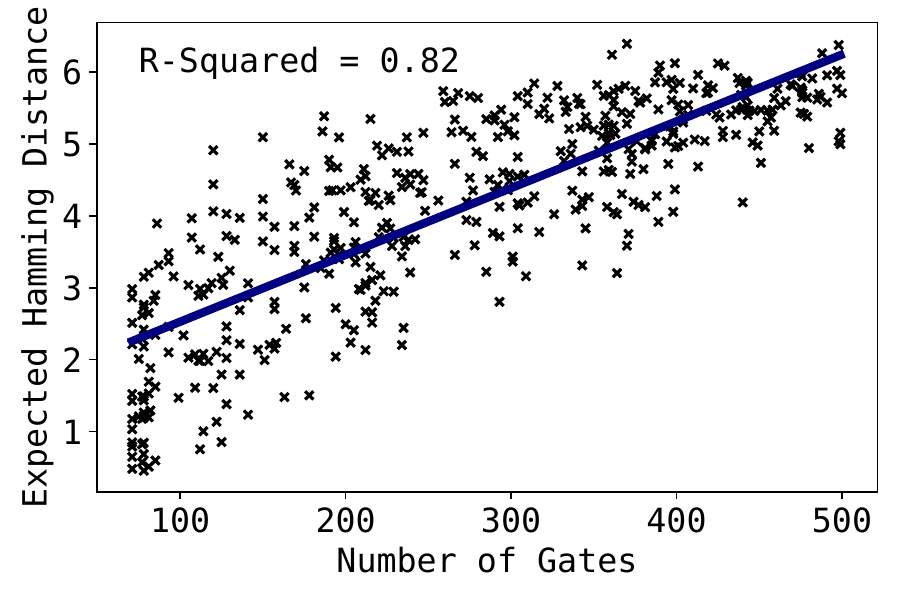}}  
    \subfigure[]{\includegraphics[width=0.263\textwidth]{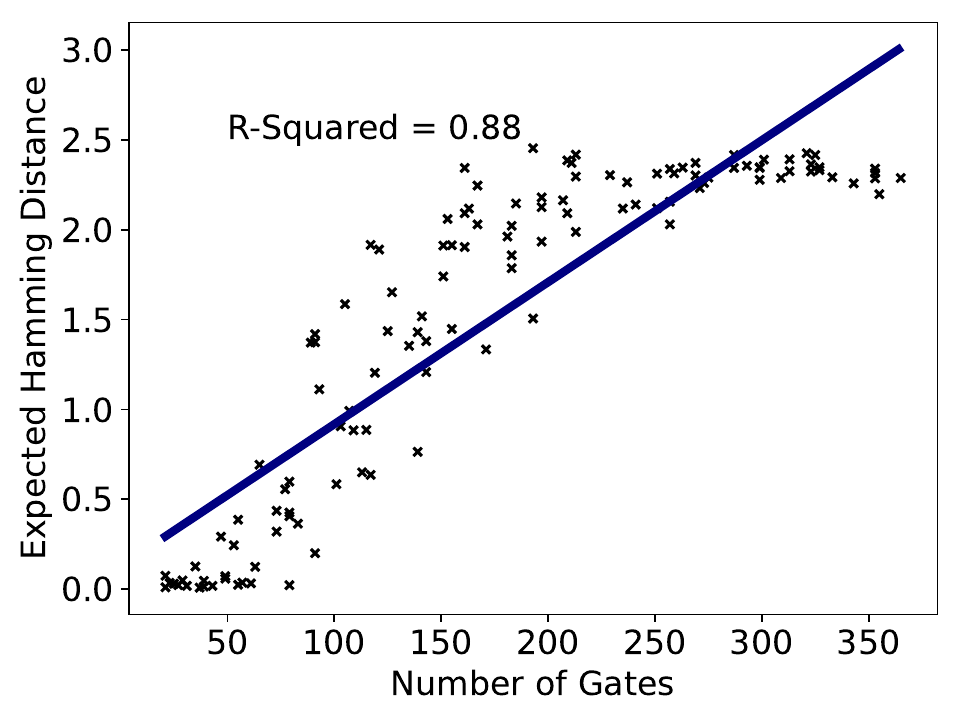}}
    \subfigure[]{\includegraphics[width=0.3\textwidth]{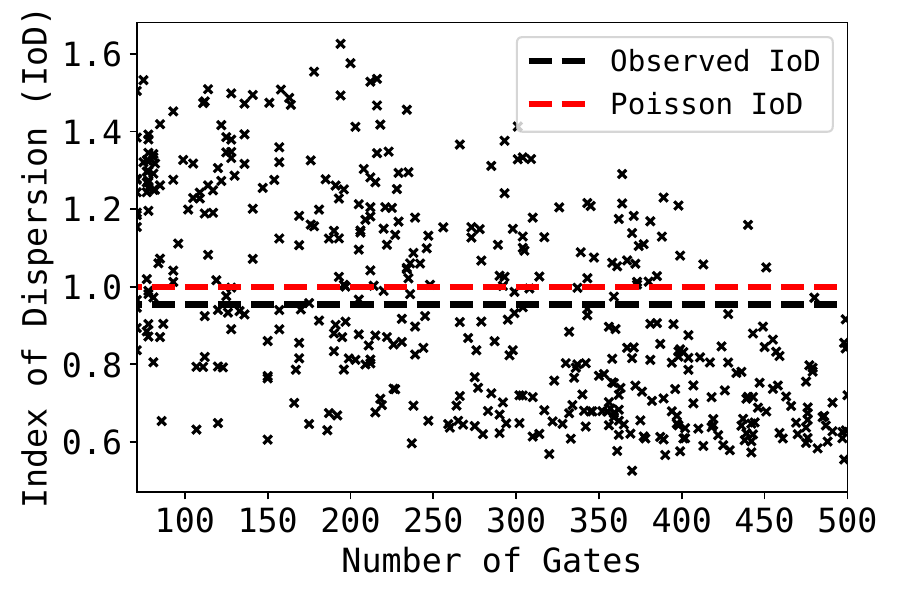}} 
    \caption{(a) Expected Hamming Distance over 500 12-qubit Randomized Benchmarking circuits over 16 IBM-Q processors (b) Expected Hamming Distance over 125 5-qubit Randomized Benchmarking circuits on IonQ's 5-qubit trapped ion processor (c) Index of Dispersion of Hamming Errors over 500 12-qubit RB Circuits results on Superconducting systems}
    \label{fig:ehd_plot}
\end{figure*}

\begin{figure*}[ht]
    \centering
    \includegraphics[width=0.8\textwidth]{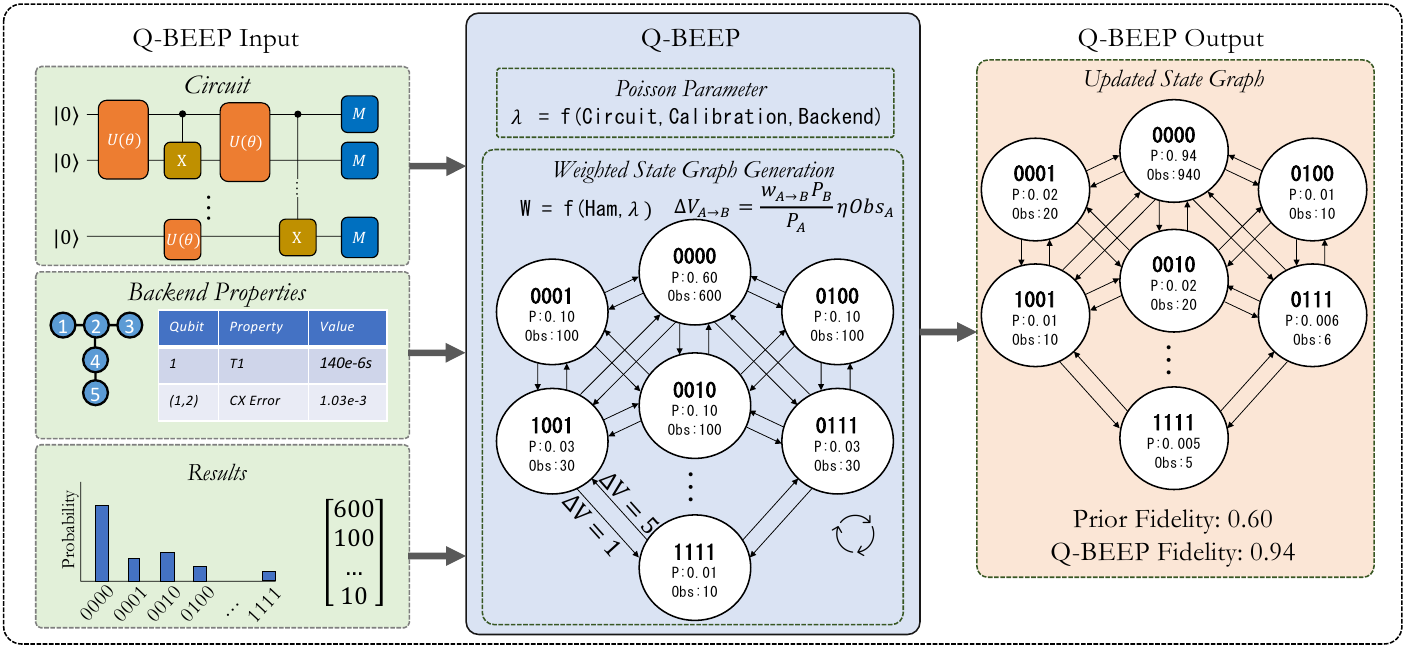}
    \caption{Overview of Q-BEEP Framework. Left box indicates the inputs comprising circuit structure induced on the backend, topology \& statistics, and resultant bit-string observations. Center block highlights the Q-BEEP structure comprising the poisson parameter estimation and the iterative state graph. The right box indicates system output with an error-mitigated state graph.}
    \label{fig:section_guide}
\end{figure*}

In this section, we delve into the empirical discovery of the clustered error structure in the Hamming spectrum, outline the design process of \sol{}, discuss the structure of the Hamming spectrum model, and explore the underlying motivation behind its corrective algorithm. We demonstrate the \sol{} architecture in Figure \ref{fig:section_guide}. \sol{}'s input comprises the induced circuit structure, backend statistics, and results. These feed into \sol{}, whereby the circuit structure and backend statistics are used to estimate the latent Hamming structure. This estimated structure is used in the iterative state graph adjustment whereby errors are mitigated. Finally, an error mitigated distribution is returned. The algorithm is outlined in Algorithm \ref{alg:master}.

\subsection{Clustered Hamming Errors at a Distance}
The structure of the Hamming spectrum described in HAMMER \cite{tannu2022hammer} empirically shows that errors in this space are clustered locally, due to the observation that the expected Hamming distance (EHD) of the observables is generally less than an EHD of $n/2$. 

We empirically discover that there exists a structure between these two - namely, non-locally clustered Hamming errors. As seen in Figure~\ref{fig:sample_distributions} we discover that there exists a large number of cases between these two, whereby Hamming errors cluster at a distance, empirically demonstrated in Figures \ref{fig:sample_distributions} and \ref{fig:ehd_plot}. The exact reasoning and intuition behind this latent structure is theoretically challenging to analyze as quantum noise channels arise from multiple sources, some of which are unknown. Notably, we do not observe this non-local clustering phenomena on noisy simulation of 13-15 qubit systems simulating the Bernstein-Vazirani algorithm, with system properties sourced from \cite{acharya2022suppressing}, indicating this phenomena is not well modelled by current noise models. Much research indicates that modeling non-Markovian noise in quantum systems is a complex task, and our understanding of the complexity behind how quantum systems behave is still lacking \cite{gao2021practical,zhang2022predicting,epstein2014investigating}. However, this does not prevent us from empirically and statistically studying and leveraging this real-world phenomenon. Our modelling of non-local errors, coupled with the iterative Bayesian State Graph error mitigation technique, represents a significant leap beyond HAMMER. Unlike HAMMER, Q-BEEP is equipped to tackle variability in machine- and circuit-specific errors.

The idea of non-locally clustered Hamming errors is depicted in Figure~\ref{fig:strucs}. Consider the unique ground truth output \texttt{1111}. In the HAMMER framework, it is expected that erroneous outputs would have a higher probability of being bit-strings with a smaller Hamming distance (e.g. \texttt{1110} with a Hamming distance of 1 and a probability of 0.3) compared to those with a larger Hamming distance (e.g. \texttt{1100} with a Hamming distance of 2 and a probability of 0.2). However, our observations have shown that this local clustering may not always hold true. As demonstrated in Figure \ref{fig:ehd_plot}, the results of more complex circuits with increased depth and decreased system performance tend to lead to an increase in cluster distance. For example, the erroneous output \texttt{0000} may occur more frequently (e.g., with a probability of 0.4) than \texttt{1110} (e.g., with a probability of 0.1). 

We empirically evaluate this on a large corpus of 12 and 5 qubit randomized benchmarking circuits using IBM's superconducting processors and IonQ's 5-qubit trapped ion processor in Figure \ref{fig:ehd_plot}. Prior to the RB circuit, we prepare a random binary state, as the homogeneous "0" state is the ground state that is naturally decayed to, and hence having a non-stable random state is better served for demonstrating the Hamming structure. In Figure \ref{fig:ehd_plot} , we compute the EHD of each circuit's real outputs, and compare them to the circuits gate count. Circuit gate count directly relates to algorithm complexity and depth, hence is a suited metric for evaluating a circuits  high level complexity. We observe a continued linear increase in the EHD of the circuit errors as circuit complexity increases on both trapped ion and superconducting systems. In conjunction with this observation, we use the metric Index of Dispersion (IoD) \cite{cox1966statistical}, defined in Equation \ref{eqn:iod}
\begin{equation}
    IoD = \frac{\sigma^2}{\mu}
    \label{eqn:iod}
\end{equation}
Whereby in Equation \ref{eqn:iod}, $\sigma^2$ is the variance, and $\mu$ is the mean. The index of dispersion is a statistical metric relating to how clustered a data set is. Increasing the index of dispersion indicates less clustering behavior, and conversely a decreasing index of dispersion indicates tighter clustering. Notably, an IoD of 1 indicates the data set is best represented by a Poisson distribution, which we will later find to best fit experimentally over a large suite of Hamming error probability distributions for multiple algorithms. We compute the IoD over each circuits Hamming spectrum, with a target bit string, and observe and average IoD of 0.92, as demonstrated in Figure \ref{fig:ehd_plot}.

To ensure that this is not a single architecture specific phenomena, we evaluate similar circuits on both trapped ion and superconducting architectures. On running 125 5-qubit randomized benchmarking circuits on IonQ's 5-qubit trapped ion processor, we compute an average IoD of 1.003, as well as a strong positive linear correlation with an $R^2$ value of 0.88. This evaluation provides empirical evidence that the structure of Hamming errors continues to be clustered for both superconducting and trapped-ion quantum devices.

This set of observations in conjunction with one and other, and the sample Bernstein-Vazirani results in Figure \ref{fig:sample_distributions}, empirically demonstrate that there does indeed exist clustering at a distance. If the mean Hamming distance continues to increase with circuit complexity, and the results remain best modelled by a Poisson distribution, the errors must remain clustered around the EHD. 

Given our understanding of the clustered nature of Hamming errors, the next step is to accurately model this non-local pattern and determine the most probable Hamming distance at which the clustering occurs. This information will be valuable in our efforts to effectively mitigate these errors.

\subsection{Modeling Hamming Spectral Errors}



Having observed the existence of non-locality in the Hamming structure of quantum algorithms, we aim to be able to correct these errors. To correct these errors however, we require a model that can estimate the post-induction Hamming error structure.  

We investigate the Hamming spectrum structure of larger algorithms, with varied degrees of entanglement, and evaluate our Hamming structure predictive model on these problems. In doing so, we analyze the Hamming Spectrum over 2750 circuits, comprising the Bernstein Vazirani algorithm, the ADDER algorithm, and randomized benchmarking circuits sampled from the Clifford group, with circuit sizes ranging from 4-15 qubits. For each of these algorithms, there is a large diversity of extremely entangled subroutines with full qubit-to-qubit communication required, to nearest neighbor qubit communication. Each of these algorithms is expected to output a unique bit-string. Hence, any output observed on a real machine that is not the expected bit-string is an error, and has a Hamming distance associated. Therefore, each solution observed on a machine has a unique Hamming spectrum.

\begin{figure}
    \centering
    \includegraphics[width=0.34\textwidth]{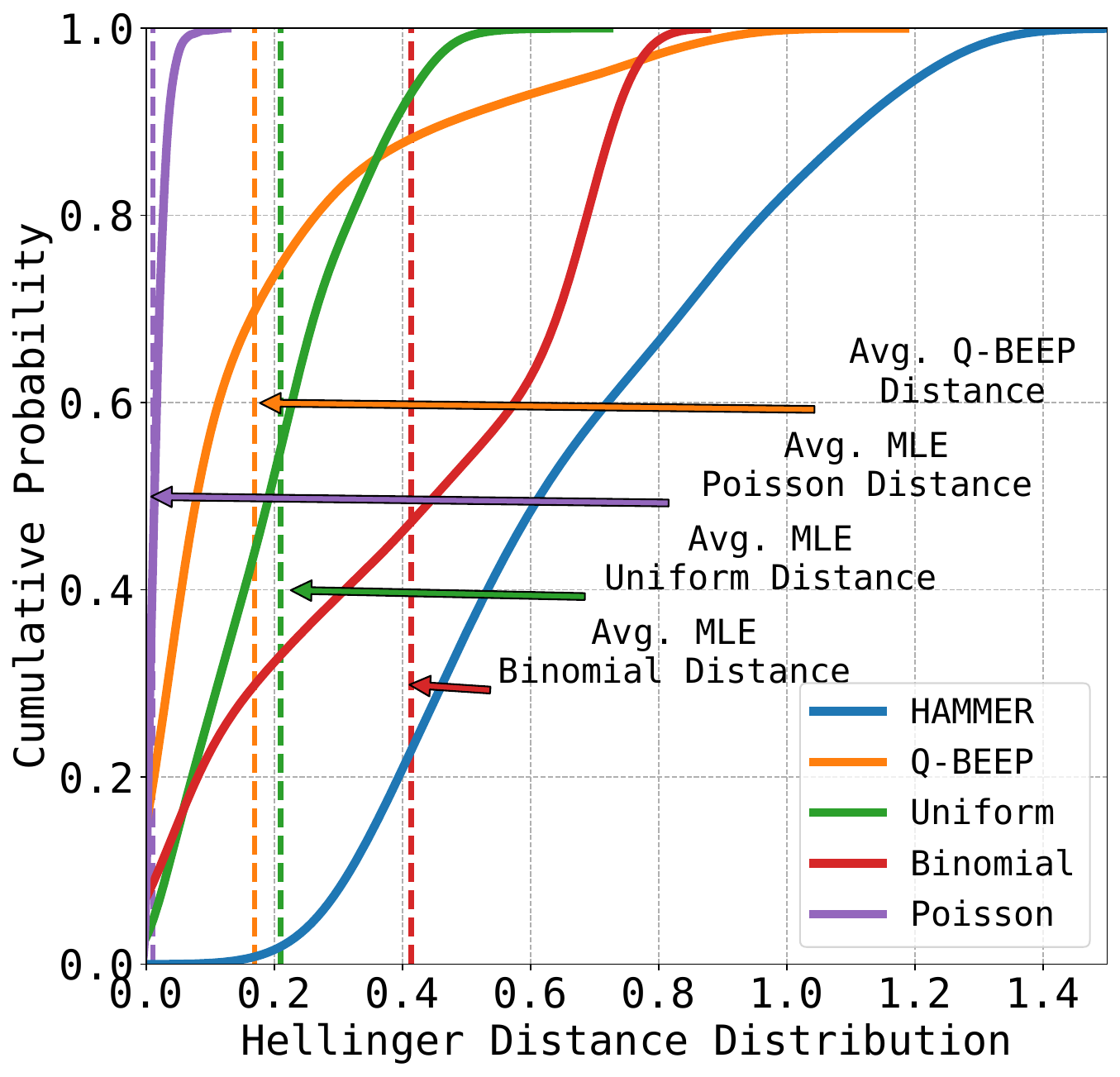}
    \caption{Validation of Q-BEEP model against 4 alternative probability distribution functions. Y-axis implies cumulative probability distribution, and X-axis Hellinger distance. Dotted vertical lines indicate average hellinger distance for model. MLE distributions are computed using the true solution and observed results, hence are the best solutions these distributions can attain. Q-BEEP is computed prior to circuit induction.}
    \label{fig:model_validation}
\end{figure}

In conjunction with this experiment, we attempt to predict the unique machine and circuit specific latent Hamming structure using a Poisson model, motivated by both the observed IoD prior, and that in Figure \ref{fig:model_validation}, the Maximum-Likelihood estimation Poisson distribution is the best performer. We model the Hamming spectrum failures as independent probabilistic failure events on a quantum processor, which are characterized by the processor calibration state and circuit structure. The relatively stable mean rate (i.e.,  $\lambda$), which describes the expected number of events occurring per time period, can be characterized by Q-BEEPs characteristic modelling Equation~\ref{eqn:poiss_param}:

\begin{equation}
    \lambda = n_Q(1-e^{\frac{-t_{circuit}}{T1}}) + n_Q(1-e^{\frac{-t_{circuit}}{T2}}) + \sum_{(i,j)}^{\sigma,\,U_\text{count}}j\sigma_i
    \label{eqn:poiss_param}
\end{equation}
where $t_{circuit}$ is the end-to-end circuit time from the pulse scheduler level; $\sigma$ is the fidelity of each respective basis gate on the processor; $n_Q$ is the number of qubits; $U_{count}$ is each respective gate count. The fidelity for each basis gate operation is reported by QC providers through benchmarking, and the respective gate operation counts are post-transpilation to processor topology and basis gate set, accounting for topological constraints and gate decomposition.  In this way, we correlate a device's statistical properties, such as gate fidelities, gate times, qubit properties, etc., as well as circuit properties, to the unique Hamming spectrum regarding a specific circuit mapping to a particular device. Using this model, we correlate as much prior information about an algorithm and its structure to the possible post-induction Hamming structure. Consequently, through Equation~\ref{eqn:poiss_param}, a Poisson distribution can be setup to model the post-induction Hamming structure of each observed bit-string. 

We examine the accuracy of the (a) Q-BEEP predictive model, which uses a Poisson distribution with an estimated rate parameter ($\lambda$) obtained through Equation~\ref{eqn:poiss_param}. We compare it to several alternative distributions, including (b) a Uniform distribution, (c) a Binomial distribution, (d) a Poisson distribution, and (e) the HAMMER weighting distribution \cite{tannu2022hammer}. To calculate the maximum likelihood parameters for each of the parameterized probability distributions (b), (c), and (d), we use the observed Hamming spectrum of the algorithm from post-induction results. In contrast, HAMMER characterizes the likelihood of bit-strings using a weighting function instead of a distribution function. It is noteworthy that (a) Q-BEEP's distribution is computed prior to circuit induction, so it has no prior knowledge of the actual output bit-strings, relying solely on circuit structure and machine calibration statistics.

Our experimental results are demonstrated in Figure \ref{fig:model_validation}. We make two key observations: (I) Compared to other distributions, the Poisson distribution with post-induction knowledge (i.e., purple curve) fits the output distribution of Hamming errors quite well, with a mean distance of merely $0.016$. This well-fit distribution to all results observed plays a key insight into \sol{}, as the Poisson distribution is an inherently clustered distribution. The ability to capture the clustered nature of Hamming spectrum errors both from short \& low entangled circuits to deep \& high degree of entanglement circuits serves as substantial motivation that this clustered Hamming spectrum error observation still exists for more complex algorithms. However, despite the Hamming spectral errors still remain clustered, the clustering is now at a distance from the true solution. Alternative distributions such as Uniform, or Binomial exhibit higher mean Hellinger distances of $0.210$ and $0.401$. (II) Based on the fact that Poisson distribution exhibits the best fit, our Q-BEEP approach of estimating the rate parameter $\lambda$ through Equation~\ref{eqn:poiss_param} without any post-induction knowledge (i.e., orange curve) also demonstrates good performance, with an average distance of $0.159$, which provides strong motivation for the model in conjunction with the low MLE Poisson average distance observation. The below-uniform value indicates that we are indeed correctly estimating the Hamming clusters, as an incorrect lambda will have little to no overlap with the true solution, whereas the uniform distribution always has some overlap. For most of the time, Q-BEEP's pre-induction model exhibits better performance than all the other non-Poisson models, including HAMMER.

In Figure~\ref{fig:sample_distributions}, we showcase 8 circuits of varying depth and width. We show the observed distribution, Q-BEEP distribution (Poisson distribution with $\lambda$ obtained through Equation~\ref{eqn:poiss_param}), and HAMMER's distribution. As can be seen, (I) with more qubits and more complex circuits (starting from 8-qubits), the observed outputs in the Hamming spectrum does not cluster locally at distance 0. Instead, under different scenarios, they cluster with different distance; (II) HAMMER's local clustering assumption prevents it from accurately modeling the observed distribution in general. Their weighting model lacks sufficient scope through the one-size-fits-all model and cannot capture the latent Hamming spectrum structure of increasingly larger and more complex circuits where the errors become localized at a distance. (III) Q-BEEP can more precisely estimate the Hamming spectrum of the output distribution by adopting the Poisson distribution model, and evaluating the circuit structure, the qubit topology, and the backend runtime statistics for the rate parameter.

\subsection{Bayesian Reclassification of Bit-strings}

Described the error model of \sol{}, we now move onto motivating how we can correct and mitigate these errors. Note that each observational output has a probability of being the correct output, and a probability of being an incorrectly classified result. The latter probability is characterised by the Bayesian inference of the equation $P(Class = BS-Actual | BS-Observed)$, where we are trying to discover the probability of an observed bit-string (BS) belonging to the correct solution of BS-Actual. Each observed result has a probability of belonging to another class. We motivate the utilization of the latent Hamming structure to re-frame this model as $P(Class = BS-Ham(n) | Observed BS)$, where $BS-Ham(n)$ defines a bit-string of Hamming distance $n$ from the observed bit-string. Using Bayes theorem, we can reframe our problem through Equation~\ref{eqn:bayes}:
\begin{equation}
\begin{split}
    P(BS-Ham(n) | Observed BS) = \\ \frac{P(Observed BS|BS-Ham(n))P(BS-Ham(n))}{P(Observed BS)}
    \label{eqn:bayes}
\end{split}
\end{equation}

Observing a bit-string of Hamming distance $n$ away from the true solution can be seen as a $P(Ham(n))$ failure, which is the same result generated from the Poisson model discussed. Using the discussed circuit-hardware aware Poisson model as the model for $P(Observed BS|BS-Ham(n)) = Poisson(\lambda,n)$, and by making the assumption that the observed probabilities of each bit-string are unbiased estimators of the true underlying probability, we can claim that the probability of each observed bit-string belongs to another bit-string of Hamming distance $n$ away, can be approximated by \sol{}'s characteristic equation -- Equation \ref{eqn:reclass_prob}:

\begin{equation}
\begin{split}
    P(Class = BS-Ham(n) | Observed BS) =\\ 
    \frac{\text{Poiss}(\lambda_\text{circuit},n)P(\text{BS-Ham(n))}}{P(\text{Observed BS})}
    \label{eqn:reclass_prob}
\end{split}
\end{equation}

\subsection{\sol{} Framework}

We now introduce the \sol{} framework, which is a graph state update algorithm for quantum error mitigation based on the Hamming spectrum structure of errors observed in quantum computing. \sol{} comprises a collection of observed bit-strings from the quantum algorithm, the system, the algorithm circuit, the algorithm-aware Hamming error model, and the iterative state-based update technique to reclassify erroneous results to their corrected respective bit-strings.  

\sol{} begins by modeling the Hamming spectrum of the algorithm by requesting the quantum processor characterizing statistics prior to analysis. This comprises error rates, qubit decoherence times, and topological constraints. Alongside the transpiled circuit, the Poisson parameter $\lambda$ is computed as described. Having generated the characterizing Poisson distribution of the circuit and processor, we induce a circuit and map the observed probabilities and bit-string counts to the vertices of a Bayesian network state graph. The number of vertices in this graph is equal to or less than the number of shots taken, N, on the quantum processor. Each vertex is linked by the value of $Poiss(\lambda,k)$ where Poiss refers to the poisson distribution,  and $k$ is the Hamming distance between the two vertices, at most populating r edges. To ensure scalability, an edge is established only if its weight surpasses a probability threshold $\epsilon$, set to $0.05$ in this work, providing a worst-case complexity per update of $\mathcal{O}(Nr)$.

Having established the state graph representation with edges and vertices populated, we perform state reclassification whereby each node has a probability and observation count attached to it. For every node in the graph, we iterate over each connected node and compute the portion of observations from the prior node belonging to the former node according to Equation~\ref{eqn:update_flow_eqn}:
\begin{equation}
    n_{A \xrightarrow[]{} B} = \frac{Node_{A-Obs}\times w_{Edge(A\xrightarrow[]{}B)} \times Node_{B-Prob}}{Node_{A-Prob}}
    \label{eqn:update_flow_eqn}
\end{equation}
where in Equation~\ref{eqn:update_flow_eqn}, $Obs$ refers to the observation count of the node. $Prob$ is the probability of the node. Edge weight is characterized by Equation~\ref{eqn:reclass_prob}. Once each node has its outgoing flows computed and compared with the total incoming flow and the observation count of the node. In \sol, reclassification overflow is implemented where we abide by the constraint $Node_{Inflow} + Node_{Obs} > Node_{Outflow}$. If the total outflow is greater, then a re-normalization process is applied where outflow is then updated by $Node_{Outflow} = \frac{Node_{Outflow}}{Node_{Inflow} + Node_{Obs}}$. This system reclassification procedure is executed iteratively. A learning rate is included, where a system-wide multiplication of the edge weights occurs per iteration, which scales the amount of inter-node flow. To encourage converging, and prohibit cycling between local nodes, a dampened learning rate of $1/n$ is practiced, where $n$ is the number of iterations. 

Algorithm~\ref{alg:master} shows the overall Q-BEEP algorithm. Within Algorithm~\ref{alg:master}, initial parameters are set until the first \emph{for} loop. Within the first \emph{for} loop, each graph vertex is generated with non-zero probability. In this sense, the algorithm is scalable to the number of observations taken on a quantum computer. From here, the relevant edges are generated, and to maintain classical scalability, only edges with a weight greater than $\epsilon$ are generated. $R$ is the observed results.Each vertex $V$ is iterated over in the graph $G$, with its $dV_{Out(V)}$ computed, which refers to the cumulative total outflow of vertex $V$. Each node keeps track of the amount of incoming data points $dV_{In(V)}$. Then from here, each node has its count updated according to the in-flow and out-flow of the node, and is normalized if needed. The algorithm is repeated $n$ times, where in each iteration the weights of the graph are scaled according to $E\times\eta$.
\begin{algorithm}[!t]
\caption{\sol{}}
\label{alg:master}
\begin{algorithmic}
\STATE $Require:\text{Quantum Circuit,R,n,}\eta,\epsilon,QPU_\text{Perf}$
\STATE $G \gets Graph(V=R_\text{bit-strings},E=None) $
\STATE $\lambda \gets f(Circuit,QPU_\text{perf})$
\STATE $n \gets Solve(Poisson(\lambda,n) < \epsilon) $
\FOR{BStr in R}
    \STATE $G(V=Bstr)[P] \gets P(Results=BStr) $
    \STATE $G(V=BStr)[Count] \gets Count(Results=Bstr) $
    \FOR{i in [1,2,3..n]}  
        \FOR{HammingBStr in Ham(BStr,n)}
            \STATE Init $E(BStr,HammingBStr)$ in G 
            \STATE $W_{(BStr,HammingBStr)} = Poisson(\lambda,n)$
        \ENDFOR
    \ENDFOR
\ENDFOR
\FOR{$n$}
    \STATE $E_n = E\times\eta$
    \FOR{V in G}
        \FOR{$E_n$(A-to-B) in V}
            \STATE $dV_{Out(V)} = dV_{Out(V)} + \frac{w_E \times V_B[P]}{B_A[P]}$
        \ENDFOR
    \ENDFOR
    \FOR{V in dV}

        \IF{$dV_{Out(V)}>V[Count] + dV_{In}$}
            \STATE $dV_{Out(V)} = V[Count] + dV_{In}$
        \ENDIF
        \STATE $V[Count] = V[Count] - dV_{Out(V)} + dV_{In(V)}$
    \ENDFOR
\ENDFOR
\end{algorithmic}
\end{algorithm}
\subsection{\sol{} Limitations}
\sol{} attempts to predict and exploit the latent Hamming structure of errors observed on quantum computers. This requires the errors to exhibit some structure. With respect to the predictability, \sol{} uses a Poisson distribution modeling scheme to attempt to predict the distance of errors from the real solution, and then reconstruct the target output. When the prediction has unreliable access to system-wide information, or the $\lambda$-estimation is substantially incorrect, \sol{} will struggle to perform error mitigation. This is also shown in Figure~\ref{fig:model_validation}, where \sol{} beats out uniform distribution until the 84th percentile. When \sol{} inaccurately estimates the Poisson parameter, it can seek to correct from a pool that has lower support. This can be thought of as two Poisson distributions of substantially different $\lambda$s, which leads to low overlap, compared to the uniform distribution, which always has some overlap. Finally, it is required that the probability of the correct bit-string being non-zero, as in the initialization phase, only vertices in the state graph that have non-zero probabilities are generated. Hence, if the true solution was never observed, it will not be correctly targeted by Q-BEEP. Q-BEEP is scalable until algorithms become predominantly noise, which is the case for most algorithms in the near term. However, as hardware and software continue to improve, the reach of Q-BEEP expands. Although Q-BEEP requires a base degree of error structure, it can be used in conjunction with other error mitigation techniques like Quancorde \cite{ravi2022boosting}, which enhances the baseline fidelity from a collection of ensembles, thereby amplifying the benefits of Q-BEEP. Furthermore,  adapting and analysing how Q-BEEP interacts with QEC codes is an unexplored area, and is an interesting direction to investigate. We set this as a future work.

\section{\sol{} Evaluation}

In the following section, we will walk through \sol{}'s evaluation and comparison with the state of the art. 

\begin{figure*}[!ht]
    \centering
    \includegraphics[width=0.9\textwidth]{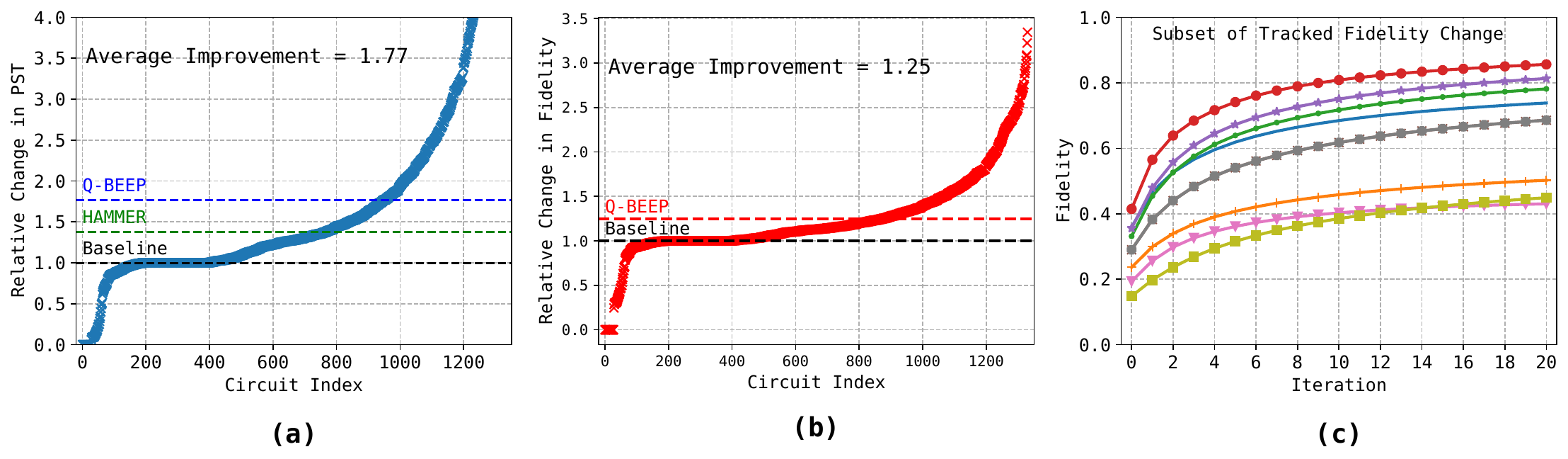}
    \caption{\sol{} Performance when applied to 165 BV circuits when transpiled to 16 machines of circuit size 5-15 qubits. \textbf{(a)} demonstrates the relative PST improvement when \sol{} is applied to Bernstein Vazirani resultant probability bit-strings, and is a direct comparison figure with HAMMER. \textbf{(b)} demonstrates the change in fidelity for the respective circuits, and \textbf{(c)} demonstrates the tracked fidelity at each state update of the problem state graph per iteration.}
    \label{fig:bv_eval}
\end{figure*}
\subsection{\sol{} Configuration and Evaluation}

Since \sol{} adopts an iterative state update scheme, it requires a prior parameter setup. For the parameter setup, we use a learning rate of $\frac{1}{\eta}$, where $\eta$ is the iteration number, and our iterations are set to 20 updates. 

As for quantum computing resources, we use a total of 16 IBMQ quantum processors, ranging in size from 5 qubits to 127 qubits. IBMQ provides daily calibration statistics for the processor performance. Furthermore, we use the Quantum Approximate Optimization Algorithm (QAOA) data set \cite{harrigan2021quantum} generated by Google on the Sycamore processor -- a 53 qubit superconducting system. Our evaluation comprises 165 BV circuits transpiled to 8 IBMQ quantum machines with varying topology, ranging in problem size from 5 to 15 qubits, 14 QASMBench circuits run on 16 IBMQ quantum machines \cite{li2020qasmbench}, and 340 QAOA results \cite{harrigan2021quantum}. We demonstrate consistent improvements in both application specific metrics, and fidelity. 

\subsection{\sol{} applied to Bernstein-Vazirani}

We begin our evaluation with the Bernstein-Vazirani algorithm, which is the primary algorithm that HAMMER benchmarks. Bernstein-Vazirani is motivated to benchmark as it is a low entropy example, with one expected output bit-string, and is easily scaled to more qubits.

\subsubsection{Evaluating BV Performance}

BV is a quantum algorithm that uncovers a hidden bit-string $s$ from the function $f(x)=s \times x\times mod(2)$. The BV algorithm hopes to produce a single bit-string output, which is the solution to the problem. Hence, when inducing on near term hardware, which produces a mixture of both the correct solution and noise, the strength of the inference can be evaluated as Probability of Successful Trial (PST). This is the ratio of correct observations to total observations, shown in Equation~\ref{eqn:pst}, where $n_\text{CorrectBitstring}$ is the number of correct observations, and $n_\text{Trials}$ is the number of shots.
\begin{equation}
    PST = {n_\text{CorrectBitstring}}/{n_\text{Trials}}
    \label{eqn:pst}
\end{equation}
Higher PST values indicate higher representation of correct bit-strings, and the quality of a circuit induction can be related to this ratio. Increasing this value is the goal of Q-BEEP. We further illustrate the fidelity, which represents the distance between the ideal and observed solutions, and the tracked fidelity of a small subset of solutions over each iteration.

\subsubsection{Bernstein-Vazirani Results}

Demonstrated in Figure \ref{fig:bv_eval}, a comprehensive evaluation of \sol{} is applied. In applying \sol{} to 165 circuits on 8 machines, we demonstrate an average PST improvement of 1.77, with up to 11.20x improvement of PST. 14.0\% of \sol{} results in a reduction in PST, which is attributed to incorrect lambda prediction. We motivate this through the notion that of the 8 quantum machines, 75\% percent of failures come from 4 machines. With respect to fidelity, we observe an average fidelity gain of 25\%, with a maximum fidelity gain of 234\%. Q-BEEP is able to operate on algorithms of limited fidelity, as seen in Figure \ref{fig:bv_eval}-(c), where Q-BEEP improved fidelity from 0.14 to 0.38. Notably, once the state approximates the maximally mixed state, i.e., when fidelity is minimized for an algorithm, there is no structure to exploit, and hence statistical error mitigation will see little to no benefit. This can be seen in Figure \ref{fig:bv_eval}-(a), where a reduction in PST occurs.

\begin{figure}
    \centering
    \includegraphics[width=0.38\textwidth]{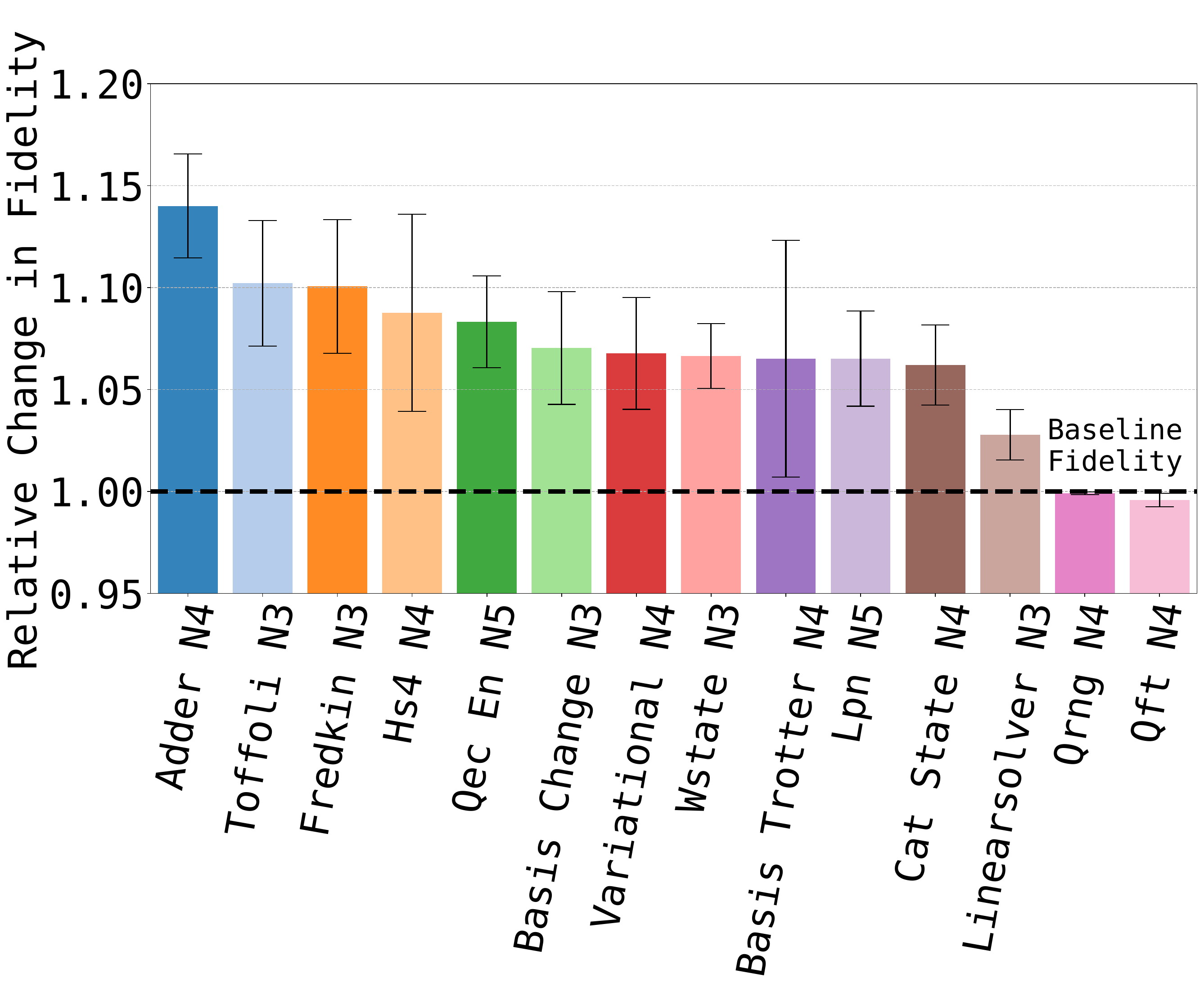}
    \caption{\sol{} performance when applied to 12 QASMBench circuits over 16 IBMQ machines. }
    \label{fig:qasmbench_perf_algo}
\end{figure}
\begin{figure}
    \centering
    \includegraphics[width=0.42\textwidth]{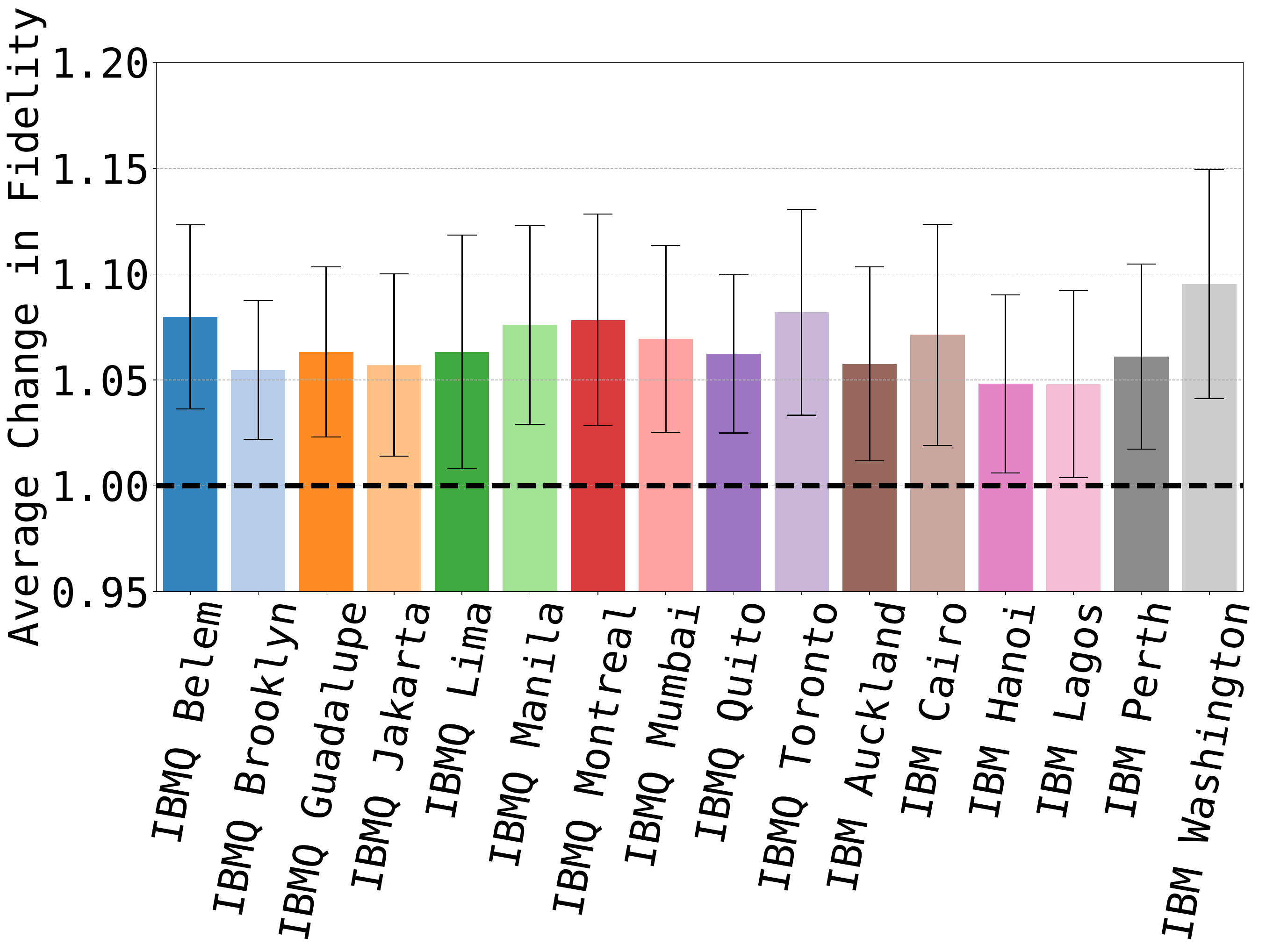}
    \caption{\sol{} performance when applied to 16 quantum machines over 12 QASMBench circuits.}
    \label{fig:qasmbench_perf_machine}
\end{figure}

\begin{figure*}[!ht]
    \centering
    \includegraphics[width=0.92\textwidth]{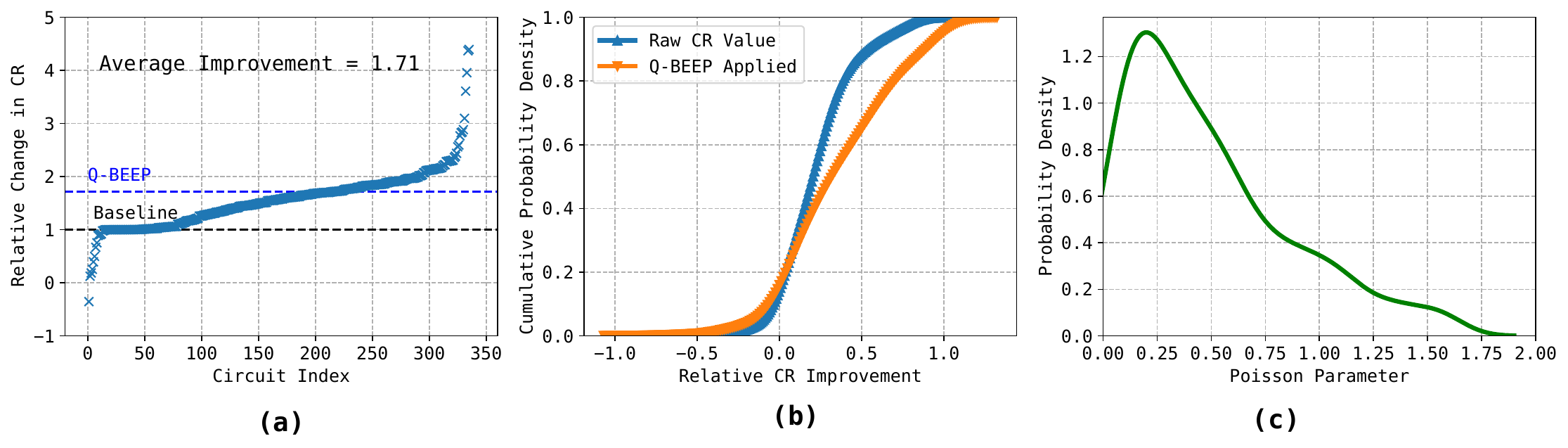}
    \caption{\sol{} Performance when applied to 340 QAOA results. \textbf{(a)} demonstrates the relative CR improvement when \sol{} is applied to QAOA resultant probability bit-strings, and is a direct comparison figure with HAMMER. \textbf{(b)} demonstrates the change in fidelity for the respective circuits , and \textbf{(c)} demonstrates the tracked fidelity at each state update of the problem state graph per iteration.}
    \label{fig:qaoa_eval}
\end{figure*}

\begin{figure}
    \centering
    \includegraphics[width=0.34\textwidth]{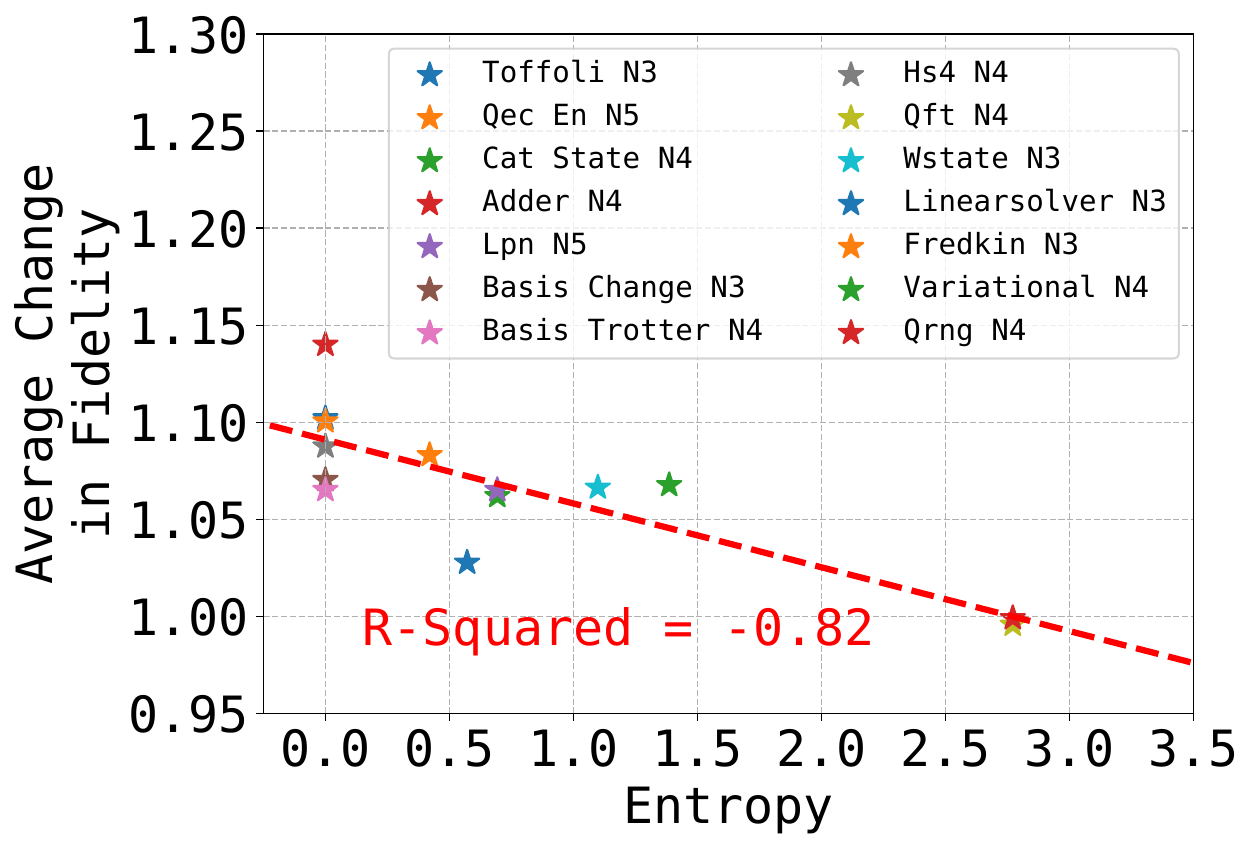}
    \caption{Entropy analysis of QASMBench algorithms and average relative fidelity improvement by \sol{}.}
    \label{fig:entropy_analysis}
\end{figure}

\subsection{\sol{} applied to QASMBench}

QASMBench is a quantum benchmarking suite \cite{li2020qasmbench} comprising a multitude of varying complexity algorithms. We benchmark Q-BEEP on QASMBench to show Q-BEEP's applicability to generalized algorithms of varying entropy.

\subsubsection{Evaluating QASMBench Performance}
QASMBench comprises a multitude of diverse algorithms. The applications described comprise high level algorithms with non-singular desired bit-string outputs. We use fidelity to compare the prior and post performance of \sol{} applied to QASMBench. 

\subsubsection{QASMBench Results}
We demonstrate the QASMBench results in Figures \ref{fig:qasmbench_perf_algo} and \ref{fig:qasmbench_perf_machine}. In Figure~\ref{fig:qasmbench_perf_algo}, we observe that the Adder algorithm attains the highest fidelity boost, with a maximum fidelity boost observed of 17.8\% on \textbf{IBM Washington}. An average performance boost across all machines and algorithms of 6.67\% is observed, including algorithms such as Qrng and Qft. Qrng and Qft are the only two algorithms that observe no performance gain, which is attributed to the nature of these algorithms. Both Qft and Qrng, when observed against the Pauli-Z measurement, will generate equal superposition of all possible bit-strings. Hence, \sol{} attempts to search for a latent Hamming structure around predominant bit-strings, and with none to find, no gain is observed. 

With respect to Figure \ref{fig:qasmbench_perf_machine}, we demonstrate that \sol{} is not machine specific, and performs across the board, with each machine evaluated demonstrating overall average fidelity improvements. This observation of consistent improvement across all machines of varying size and quality helps motivate that Q-BEEP is a general solution, and can handle diverse sets of machine performance statistics.

\subsection{\sol{} applied to QAOA}
    
\subsubsection{Evaluating QAOA Performance}

The QAOA algorithm seeks to optimize a quantum circuit such that the output state minimizes a cost function. Here, we obtain and evaluate on the raw dataset from Google \cite{harrigan2021quantum}. For all problems within this data set, they aim to minimize the expectation value of the cost function $C$. To evaluate performance, they use the Cost Ratio. Cost Ratio is the ratio of $C$ to the minimum value of $C$, or $C_\text{min}$, which is defined in Equation~\ref{eqn:cmin}:
\begin{equation}
    CR = C/C_\text{min}
    \label{eqn:cmin}
\end{equation}
Due to the fact that all problems have a negative $C_\text{min}$, improving algorithms increase their CR values. Therefore, to evaluate our performance of \sol{}, we evaluate our relative performance, $CR_\text{Improvement} = \frac{CR_\text{\sol{}}}{CR_\text{Prior}}$, where $CR_\text{\sol{}}$ is the CR post \sol{} application, and $CR_\text{Prior}$ is the CR attained in the paper.
We note that frequent calibration data of Google's 53-qubit Sycamore processor is not publicly available. Hence, we use the published statistics.

\subsubsection{QAOA Results}

As demonstrated in Figure \ref{fig:qaoa_eval}, \sol{} provides substantial improvements to QAOA. \sol{} provides a 94.1\% success rate in improving QAOA solutions, with an average improvement of 1.71$\times$. Certain solutions were boosted up to a relative CR improvement of 31.7$\times$, which are not plotted in \ref{fig:qaoa_eval}-(a) due to scaling. \ref{fig:qaoa_eval}-(b) demonstrates the overall shift in average performance by applying \sol{}. The post-application orange line shifts the S-curve of the cumulative distribution right, indicating the CR value attained increases. Finally, we demonstrate in \ref{fig:qaoa_eval}-(c) the Poisson parameter distribution. We see that the Poisson parameters for these solutions lie in the 0-2 range, and distances up to 5 Hamming distances per bit-string are evaluated.

\section{Discussion on Entropy}

We discuss entropy to help provide some insights with respect to relating Q-BEEP and its capabilities to quantum algorithm entropy. There exists high diversity in observable Shannon entropy for quantum algorithms, for example the Bernstein Vazirani has an ideal entropy of $0.0$ \cite{nagata2017generalization}, whereas Quantum Random Number generators \cite{li2020qasmbench} have a maximum entropy, with all outputs being equally likely. With quantum algorithm graph states, higher entropy's lead to a more balanced P value across each node within \sol{}'s state graph. For corrective analysis on post-induction state graphs, this creates difficulty in discerning between results that are errors and have an underlying contribution to the distribution. This can be attributed to the ratio of $\frac{P_A}{P_B}$ in \sol{}, which requires node probability imbalance across the state graph to make changes to the distribution, and to correct errors.

As demonstrated in Figure \ref{fig:entropy_analysis} we compute the entropy of each expected output distribution from QASMBench, and compare it to the performance gain. We observe a strong inverse linear correlation between the correcting ability of \sol{} and an algorithm's expected information entropy, with an $R^2$ value of -0.82. Therefore, based on this evaluation, we expect \sol{} to have better performance when applied to algorithms that have predominant outputs, and not equal probability highly diverse algorithms.

\section{Related Work}

Reducing quantum errors is crucial to the success of quantum computing. The field of handling quantum errors within the system and architecture community is rapidly growing with two predominant techniques: quantum error correction (QEC) and quantum error mitigation (QEM). 


\textbf{Quantum error correction} currently is predominantly focused on building surface codes \cite{fowler2009high,devitt2013quantum,lidar2013quantum,gottesman1997stabilizer}. Surface codes employ the idea of a logical qubit, which is comprised of multiple physical qubits, all of which are responsible for correcting quantum errors. Surface codes utilize ancilla qubits entangled which are measured during circuit induction, and error correcting operations are applied to the logical qubit such that the error is corrected . However, surface codes require errors to be corrected at a rate faster than they are generated. Wu et al. \cite{wu2022synthesis} describes a procedure for stitching surface codes to near term superconducting quantum computers, which tackles the mapping problem of syndrome and ancilla qubits to superconducting devices based on degrees of connectivity. LILIPUT \cite{das2022lilliput} proposes a lightweight lookup table for syndrome error correction, motivated by the need for live rapid syndrome error correction. QULATIS \cite{ueno2022qulatis} discusses a syndrome decoder design capable of operating within a cryogenic environment, due to the expected operating conditions of fault tolerant quantum computing.

\textbf{Quantum Error Mitigation} comprises pre and post circuit induction error mitigation techniques \cite{lowe2021unified,kim2020quantum,lao20222qan,alam2020circuit,patel2021qraft}. Pre-circuit induction is predominantly characterised through transpiler improvements. Due to the fact that quantum errors are generated via increasing circuit depth and gate counts, therefore optimizing to minimize these features is advantageous to system performance. Gushu et al.~\cite{li2019tackling} discusses a transpilation technique comprising graph search techniques and an optimization function. Application specific transpiation techniques \cite{alam2020circuit,lao20222qan,gokhale2019partial} seek to apply gate minimization and cancellation techniques by exploiting domain specific knowledge. For example, 2QAN \cite{lao20222qan} is a transpilation technique applied to 2-local Hamiltonian simulation that operates by exploiting the sequencing invariance of the Hamiltonian term operators, and hence attempts to place operations that can be cancelled and optimized closer to each other. As for post-circuit induction quantum error mitigation techniques, Zheng et al. \cite{zheng2020bayesian} uses a Bayesian inference algorithm to identify posterior distributions to mitigate post-induction errors. Patel et. al. \cite{patel2021qraft} uses the reversibility of quantum circuits to mitigate post-induction errors via reversed circuit re-induction. Hamming spectrum error mitigation is dominated by HAMMER \cite{tannu2022hammer}, which is a pioneering work in combining the structure of errors in quantum computers along side a reclassification protocol to boost circuit fidelities.
\section{Conclusion}

In this paper, we present \sol{}, a system comprising a characterizing Hamming spectrum model, capable of modeling both localized and distant clustered Hamming errors. \sol{} uses this model and an iterative approach over a Bayesian network to perform highly performant quantum error mitigation. \sol{} is comprehensively evaluated on BV, QASMBench, and QAOA, gaining up to $234.6\%$ fidelity improvements. \sol{} provides more insights into the latent Hamming error structure, and generate a lightweight offline QEM model that requires no modification to the circuit or the machine. \sol{} motivates further research in the Hamming spectral quantum error mitigation domain, with potential further investigation into a better $\lambda$ estimation function or better Hamming spectrum characterization equations.

\section*{Acknowledgements}
This material is based upon work supported by the U.S. Department of Energy, Office of Science, National Quantum Information Science Research Centers, Co-design Center for Quantum Advantage ($C^2QA$) under contract number DESC0012704. Yufei Ding was supported by NSF 2048144 and would like to acknowledge the support from the Robert N. Noyce Trust. We would like to thank the PNNL operated IBM-Q Hub. This research used resources of the National Energy Research Scientific Computing Center (NERSC), a U.S. Department of Energy Office of Science User Facility located at Lawrence Berkeley National Laboratory, operated under Contract No. DE-AC02-05CH11231. This research used resources of the Oak Ridge Leadership Computing Facility, which is a DOE Office of Science User Facility supported under Contract DE-AC05-00OR22725. The Pacific Northwest National Laboratory is operated by Battelle for the U.S. Department of Energy under Contract DE-AC05-76RL01830.

\newpage
\bibliographystyle{IEEEtranS}
\bibliography{refs}

\end{document}